\documentclass[11pt]{article}
\usepackage{amsmath}
\usepackage{amssymb}
\usepackage{graphicx}
\usepackage{cite}
\usepackage{color} 
\usepackage[labelfont=bf,labelsep=period,justification=raggedright]{caption}
\makeatletter
\renewcommand{\@biblabel}[1]{\quad#1.}

\makeatother
\date{}
\begin{document}
\begin{flushleft}
{\Large
\textbf{Avalanches in self-organized critical neural networks:\newline A minimal model for the neural SOC universality class}\footnote{published as: PLoS ONE 9(4): e93090 (2014)}
} 
\medskip \\
{\small
Matthias Rybarsch$^{1}$, 
Stefan Bornholdt$^{1,\dagger}$
\medskip \\
\bf{$^1$} Institut f\"{u}r Theoretische Physik, Universit\"at Bremen, D-28359 Bremen, Germany 
\\
$\dagger$ E-mail: bornholdt@itp.uni-bremen.de 
}
\end{flushleft} 

\section*{Abstract} 
The brain keeps its overall dynamics in a corridor of intermediate activity 
and it has been a long standing question what possible mechanism could 
achieve this task. Mechanisms from the field of statistical physics 
have long been suggesting that this homeostasis of brain activity could 
occur even without a central regulator, via self-organization on the level 
of neurons and their interactions, alone. Such physical mechanisms 
from the class of self-organized criticality exhibit characteristic 
dynamical signatures, similar to seismic activity related to earthquakes. 
Measurements of cortex rest activity showed first signs of dynamical 
signatures potentially pointing to self-organized critical dynamics 
in the brain. Indeed, recent more accurate measurements allowed for a 
detailed comparison with scaling theory of non-equilibrium critical 
phenomena, proving the existence of criticality in cortex dynamics. 
We here compare this new evaluation of cortex activity data to the 
predictions of the earliest physics spin model of self-organized 
critical neural networks. We find that the model matches with the 
recent experimental data and its interpretation in terms of dynamical 
signatures for criticality in the brain. 
The combination of signatures for criticality, power law distributions of 
avalanche sizes and durations, as well as a specific scaling relationship
between anomalous exponents, defines a universality class characteristic 
of the particular critical phenomenon observed in the neural experiments. 
The spin model is a candidate for a minimal model of a self-organized 
critical adaptive network for the universality class of neural criticality.  
As a prototype model, it provides the background for models that include 
more biological details, yet share the same universality class 
characteristic of the homeostasis of activity in the brain. 

\section*{Introduction}

Information processing by a network of dynamical elements is a delicate matter: 
Avalanches of activity can die out if the network is not connected enough or 
if the elements are not sensitive enough; on the other hand, 
activity avalanches can grow and spread over the entire network 
and override information processing, as e.g.\ observed in epilepsy. 
Therefore, it has long been argued that neural networks have to 
establish and maintain a certain intermediate level of activity 
in order to keep away from, both, the regimes of chaos and silence
\cite{Langton:1990,Herz:1995,Bak:2001,Bornholdt:2003}. 
Similar ideas were also formulated in the context of genetic networks 
where Kauffman postulated that information processing in these evolved 
biochemical networks would be optimal near the ``edge of chaos'', 
or the critical regime of the dynamical percolation transition of 
such networks \cite{Kauffman:1993}. 

In the wake of the discovery of self-organized criticality (SOC) it was asked 
if also neural systems were self-organized to some form of criticality 
\cite{Bak:1988}. An early example of a SOC model that had been adapted to be 
applicable to neural networks is the model by Eurich et al. \cite{Eurich:2002}.
Their model is a variant of the random neighbor Olami-Feder-Christensen 
model for earthquakes and exhibits, subject to one critical coupling parameter, 
distributions of avalanche sizes and durations which they postulate could 
also occur in neural systems. 

Another early example is a spin model for self-organized critical neural networks 
\cite{Bornholdt:2001,Bornholdt:2003} that draws on the alternative approach 
of self-organized critical adaptive networks \cite{Bornholdt:2000a}. 
Here networks are able to self-regulate towards and maintain a critical 
system state, via simple local rewiring rules which are plausible in the 
biological context.

Only after these first hypothetical models, experimental evidence for criticality 
in neural systems has been found in terms of spatio-temporal activity 
avalanches, first in the seminal work of Beggs and Plenz \cite{Beggs:2003}. 
Much further experimental evidence has been collected since, which we will 
briefly review below. Only recently, however, experimental data has 
reached the resolution to discuss the hypothesis of dynamical criticality 
in neural tissue in the context of measurements. A major finding is 
that these new data match well with scaling theory of non-equilibrium 
critical phenomena, providing us with a solid evidence for criticality in 
cortex tissue dynamics \cite{Friedman:2012}. 
As a result this sheds new light on the early spin models of self-organized 
critical adaptive neural networks, where now their predictions can actually 
be tested against the new observations. This is the purpose of this paper. 

The outline of this paper is as follows. We will first briefly review 
the further experiments on neural activity avalanches. Then we will give
an overview of models that have been motivated by these observations. 
We will then revisit the earliest spin model for self-organized critical 
adaptive neural networks \cite{Bornholdt:2003} and test its applicability 
in the light of experimental data. We redefine the model for a natural 
representation of activity avalanches \cite{Rybarsch:2010}, study the
avalanche dynamics of the model, and discuss its relation to criticality 
in the context of the scaling theory of non-equilibrium critical phenomena. 

\subsection*{Avalanche dynamics in neuronal systems}
Let us first briefly review the experimental studies on neuronal avalanche 
dynamics. In 2003, Beggs and Plenz published their findings about a novel 
mode of activity in neocortical neuron circuits \cite{Beggs:2003}. 
During \emph{in-vitro} experiments with cortex slice cultures of the rat, 
they found evidence of spontaneous bursts and avalanche-like propagation 
of activity followed by silent periods of various lengths. 
The observed power-law distribution of event sizes indicates that the 
neuronal network is maintained in a critical state. 
Also, the spatio-temporal patterns of the avalanches are stable and precise 
over many hours and robust against external perturbations \cite{Beggs:2004}, 
which indicates that they might play a central role for brain functions 
as, for example, information storage and processing. 
Neuronal avalanches have also been found during developmental stages 
of \emph{in-vitro} cortex slice cultures from newborn rats \cite{Stewart:2008}, 
as well as in cultures of dissociated neurons in different kinds of networks, 
as rat hippocampal neurons and leech ganglia \cite{Mazzoni:2007},
or rat embryos \cite{Pasquale:2008}.

Aside from these \emph{in-vitro} experiments, extensive studies \emph{in-vivo} 
have since been conducted. The emergence of spontaneous neuronal 
avalanches has been shown in anaesthesized rats during cortical development 
\cite{Gireesh:2008} as well as in awake rhesus monkeys during ongoing 
cortical synchronization \cite{Petermann:2009}.

The biological relevance of the avalanche-like propagation of activity in 
conjunction with a critical state of the neuronal network has been 
emphasized in several works recently. Such network activity has proven 
to be optimal for maximum dynamical range \cite{Shew:2009,Kinouchi:2006}, 
maximal information capacity and transmission capability \cite{Shew:2011}, 
as well as for a maximal variability of phase synchronization \cite{Yang:2012}.
Most recently, experimental evidence for universality of critical dynamics 
has been found in neuronal avalanche data \cite{Friedman:2012,Priesemann:2013,Scarpetta:2013}
and formally linked to universal scaling theory \cite{Sethna:2001}. This 
can be considered as providing a solid evidence for dynamical criticality 
in neuronal systems. 

\subsection*{Models for neural criticality}

These experimental studies with their rich phenomenology sparked 
a large number of theoretical studies and models for criticality and 
self-organization in neural networks, ranging from simple toy models 
to detailed representations of biological functions. 

A variety of models have been constructed that are careful to include 
biological details at the neuron level as a basis for possible 
self-organization. Such mechanisms include threshold firing dynamics 
and activity-dependent plasticity of synaptic couplings as the basis 
for self-organization. While some models feature synaptic facilitation 
following a firing event \cite{Arcangelis:2006,Pellegrini:2007,Meisel:2009}, 
others use synaptic depression as the main driving force towards 
criticality \cite{Levina:2007,Levina:2009}. 
It has been shown that anti-Hebbian evolution is generally capable 
of creating a dynamically critical network when the anti-Hebbian 
rule affects only symmetric components of the connectivity matrix, 
while the anti-symmetric component remains as an independent 
degree of freedom utilizable for e.g.\ learning tasks \cite{Magnasco:2009}. 
Also, synaptic plasticity on two different timescales has been discussed 
\cite{Peng:2013}. 

On the other hand, the biological plausibility of activity-dependent synaptic 
plasticity for adaptive self-organized critical networks has been 
emphasized \cite{Droste:2012}. Recently, correlations of subsequent 
firing events again came into focus as a synaptic facilitation criterion 
\cite{Arcangelis:2012}. The biological relevance of the critical state 
in neural networks for a brain function as learning has further been 
underlined \cite{Arcangelis:2010}. Most recently, the temporal 
organization of neuronal avalanches in real cortical networks has 
been linked to the existence of alternating states of high vs.\ low 
activity in the network as well as to a balance of excitation and 
inhibition in a critical network \cite{Lombardi:2012}.

While the proposed organizational mechanisms strongly differ between 
the individual models, there are signs that the resulting evolved 
critical networks may be part of the same fundamental universality 
class. Many of the models exhibit at least some of the avalanche 
statistics seen in the experimental data, as e.g.\ a power-law 
distribution with exponent around $-3/2$ for the distribution 
of avalanche sizes. 
With the recent, more detailed models in mind, we are especially 
interested in the underlying universality of self-organization.

\subsection*{Revisiting the spin model of self-organized critical neural networks}
Let us now revisit the earliest spin models of self-organized critical 
neural networks \cite{Bornholdt:2000a, Bornholdt:2003} in a formulation 
that allows for studying its avalanche dynamics in time and space. 
Two main aspects have to be addressed. 

First, the spin-type description of the dynamical variables, due to 
its symmetrized nature, does not allow to sample avalanche statistics 
at the critical point. We therefore translate the model into a version 
with Boolean state nodes and redefine its activation threshold function 
and its network rewiring mechanism accordingly. As a result, 
activity avalanches intrinsically occur in the network, 
whereas spin networks typically exhibit continuous fluctuations 
with no avalanches directly visible. 
The further advantages of this transformation in the context of biological 
networks have been discussed in a previous paper \cite{Rybarsch:2010}.

The second aspect to be reviewed is the topology the algorithm operates on. 
While the original correlation-based rewiring mechanism of network 
self-organization \cite{Bornholdt:2003} has been defined to simply operate 
on neighboring nodes on a lattice, we would like to study the model here
as an arbitrary self-organizing network, without specifying any underlying 
topology. However, while on a lattice the number of possible neighbors of 
a node is strictly limited, on a large random network near critical 
connectivity there are far more unconnected pairs of nodes than there 
are connected pairs. Thus, randomly selecting pairs of nodes for rewiring 
would introduce a strong bias towards connecting nodes which were 
previously unconnected. This bias would result in a strong increase of 
connectivity, far above any self-organized critical regime. 
Consequently, we will adapt the rewiring mechanism below to include 
arbitrary topologies without such a bias.  

The philosophy of the model is its capability of self-regulation towards a 
critical state despite being simplified to the most minimal model possible. 
Its rewiring mechanism is based on a simple rewiring rule, which only uses 
information accessible to individual nodes locally, which here means  
pre-\ and post-synaptic activities of the particular node, as well as 
correlations of these activities. 
 
\section*{Methods}
\subsection*{Adaptive network evolution}
We will now first define the dynamics {\sl on} the network and will then 
proceed with the rewiring dynamics, i.e.\ the dynamics {\sl of} the network. 

Consider a randomly connected network of $N$ nodes of Boolean states 
$\sigma_i \in \{ 0, 1 \}$ which can be linked by asymmetric directed 
couplings $c_{ij} = \pm 1$. Node pairs which are not linked have their 
coupling set to $c_{ij} = 0$. Links may exist between any two nodes, 
so there is no underlying spatial topology in this model. 
Let $K$ denote the average connectivity of the network, 
i.e.\ the number of in-links averaged over all $N$ nodes.

All nodes are updated synchronously in discrete time steps via a simple 
threshold function of their input signals with a small thermal noise 
introduced by the inverse temperature $\beta$ in the same way as in 
the original version of the model \cite{Bornholdt:2003}. However, now 
an input shift of $-0.5$ adds to the Glauber update, representing the 
modified update function in the course of the transition from spins 
to Boolean node values \cite{Rybarsch:2010}: 
\begin{eqnarray}\label{eq:01-update}
	\mathrm{Prob} [ \sigma_i(t+1) = 1 ] &=& g_\beta(f_i(t)) \nonumber \\
	\mathrm{Prob} [ \sigma_i(t+1) = 0 ] &=& 1- g_\beta(f_i(t))
\end{eqnarray} 
with
\begin{equation}
	f_i(t) = \sum_{j=1}^N c_{ij} \sigma_j(t) - \Theta_i 
\end{equation}
and
\begin{equation}
	g_\beta(f_i(t))=\frac{1}{1+\exp(-2\beta (f_i(t)-0.5))}.
\end{equation}
For simplicity, we assume that all nodes have an identical activation 
threshold of $\Theta_i = 0$, unless stated otherwise.

\subsection*{Rewiring algorithm}
The correlation-based rewiring mechanism of the original model 
\cite{Bornholdt:2003} has to be carefully revised as well, when 
changing from spin variables to Boolean variables, as inactive 
nodes are now represented by a value of $0$ instead of $-1$ 
which affects the calculation of correlations. 

The adaptation algorithm operates as follows. 
After initializing the network with random links at a given initial 
connectivity $K_{ini}$ and initial states set to $0$, 
all nodes are synchronously updated in parallel according to 
eq.\ \eqref{eq:01-update}. All activity then observed in this model 
originates from small perturbations by thermal noise, leading to 
activity avalanches of various sizes. In the following we set the 
inverse temperature to $\beta=5$. On a larger time scale, here 
after $\tau=100$ updates, a rewiring is introduced at 
one randomly chosen, single node. The new element in our revised 
model is to test whether the addition or removal of one random 
in-link at the selected node will increase the average dynamical 
correlation to all existing inputs of that node. By selecting 
only one single node for this procedure, we effectively diminish 
the bias of selecting mostly unconnected node pairs -- but 
retain the biologically inspired idea for a Hebbian, 
correlation-based rewiring mechanism on the basis of locally 
available information, only. 

Now, we have to define what is meant by \emph{dynamical correlation} 
in this case. We here use the Pearson correlation coefficient to 
first determine the correlation between a node $i$ and one of its 
inputs $j$ over the preceding $\tau$ time steps:
\begin{equation}\label{eq:PearsonCorrelation}
	C_{ij} = \frac{\langle \sigma_i(t+1) \sigma_j(t) \rangle - 
	\langle \sigma_i(t+1) \rangle \langle \sigma_j(t) \rangle}{S_i \cdot S_j}
\end{equation}
where $S_{i}$ and $S_{j}$ in the denominator denote the standard 
deviations of the states of nodes $i$ and $j$ respectively. 
In case one or both of the nodes remain frozen in their state 
(i.e.\ yield a standard deviation of 0), we will assume a correlation 
of $C_{ij}=0$, as otherwise the Pearson correlation coefficient would 
not be well defined.   

Note that we always use the state of node $i$ at one time step later 
than node $j$, thereby taking into account the signal transmission 
time of one time step from one node to the next one. Finally, 
we define the average input correlation $C_i^{avg}$ of node $i$ as
\begin{equation}\label{eq:AverageInputCorrelation}
	C_i^{avg} = \frac{1}{k_i} \sum_{j=0}^N \vert c_{ij} \vert C_{ij}
\end{equation}
where $k_i$ is the in-degree of node $i$. The factor $\vert c_{ij} \vert$ 
ensures that correlations are only measured where links are present 
between the nodes. 
For nodes without any in-links ($k_i = 0$) we define $C_i^{avg} := 0$.

In detail, the adaptive rewiring is now performed in the following steps:
\begin{enumerate}
	\item Select a random node $i$ at which the next rewiring will take place.
	\item Run network updates for $\tau=100$ simulation time steps and measure the average input correlation $C_i^{avg}$ of node $i$.
	\item With equal probability, either
		\begin{enumerate}
			\item insert an additional in-link of random weight $c_{ix} = \pm 1$ at node $i$ from a random, previously unlinked node $x$, or
			\item remove one of the existing in-links at node $i$.
		\end{enumerate}
	\item Again run $\tau=100$ network updates and measure the new $C_i^{avg}$ of node $i$ after the local rewiring at this node.
	\item If $C_i^{avg}$ has increased after the insertion or removal of the in-link, the rewiring  from step 3 is retained; otherwise, it is reverted.
	\item Run $\tau=100$ network updates to allow for a transient period prior to the next rewiring process. Iterate from step 1.
\end{enumerate}
Note that the exact choice of $\tau$ is not critical, but is chosen as $\tau=100$ 
here to ensure time scale separation of at least two orders of magnitude between 
node dynamics (fast) and rewiring changes (slow).

It is also worth noting that this updated model version -- 
in the same way as the original model \cite{Bornholdt:2003} -- 
is solely based on locally available information at synapse level 
and takes into account both pre- and post-synaptic activity. 
This is a fundamental difference to approaches discussed 
e.g.\ in \cite{Arcangelis:2006}, \cite{Pellegrini:2007} 
or \cite{Levina:2007}, where only pre-synaptic activity 
determines changes to the coupling weights.

In order to obtain an indication of the current dynamical regime 
of the network (i.e.\ whether the network is sub- or super-critical, 
or close to the critical point), we continuously measure a branching 
parameter based on potential damage spreading in the network. 
This is realized by counting, for each individual node $i$, 
the number of descendant nodes which would possibly change 
their states at time step $t+1$ if the state of node $i$ was 
changed at the present time $t$. Here, both, the present states 
(on or off) of node $i$ and its descendants, as well as the nature 
of their respective links (activating or inhibiting) are taken 
into account. The obtained number of descendant nodes prone to 
damage spreading (and thus also to signal propagation) is then 
averaged over the entire network, resulting in the branching 
parameter $\lambda$. This allows to estimate (based on the 
current network configuration) whether the network is sub-critical 
\mbox{($\lambda < 1$)} or super-critical \mbox{($\lambda > 1$)}.
For the analysis of avalanche statistics in the 
evolved critical networks we export snapshots of the network 
structure whenever the branching parameter is close to one 
(here when \mbox{$\vert \lambda - 1 \vert \leq 0.01$}).

\subsection*{Avalanche analysis}
For a detailed analysis of avalanche statistics, we use the snapshots 
of near-critical (per branching parameter estimation) network structures 
from the adaptation runs as outlined above. 
Key observables are the avalanche size $S$, i.e.\ the total number 
of nodes which become active at least once during one avalanche, 
and the avalanche duration $T$, i.e.\ the number of simulation time 
steps from the start (first node active) to the termination 
(no more nodes active) of an avalanche. To obtain those, the network 
is now run in a deterministic mode with $\beta \rightarrow \infty$. 
The update function from \eqref{eq:01-update} then simplifies to
\begin{equation}
	\sigma_i(t+1) = \Theta_0(f_i(t)) 
	\label{eq:SignumFunction}
\end{equation}
with a redefined threshold function (nodes only become active with 
activating, non-zero input)
\begin{equation} 
\Theta_0(x)= 
\left\{ 
\begin{array}{c} 
1, \ \ \  x>0 \noindent \\  
0, \ \ \ x \leq 0  
\end{array} 
\right. 
\end{equation}  
and the usual input function
\begin{equation}
	f_i(t) = \sum_{j=1}^N c_{ij} \sigma_j(t) - \theta_i
	\label{eq:InputSum}
\end{equation}
where activation thresholds $\theta_i$ are set to $0$ for now. 
With parallel updates, any network activity would eventually 
end up in either a fixed point or limit cycle attractor, 
but not necessarily at the fixed point ``all nodes off'', 
terminating an avalanche. Therefore, we introduce an exhaust 
time parameter $\Omega$ which can be biologically interpreted 
as an effect of depleting neuro-transmitters at active synapses. 
In each time step, every node will increase its individual activation 
threshold $\theta_i$ to 1 with a probability corresponding to 
its own average activity over the last $\Omega$ time steps (i.e.\ 
number of time steps where $\sigma_i=1$ divided by the total 
number of time steps $\Omega$). This turns out to be sufficient 
to eventually step out of a periodic attractor and terminate the 
avalanche. Whenever one avalanche is terminated (all nodes off), 
we will start a new one by randomly activating one single node 
and continue with the parallel updates. 
We constantly keep track of cumulative avalanche size and duration 
distributions, $f_{cml}(S)$ and $f_{cml}(T)$, as well the average 
size $\langle S \rangle (T)$ of avalanches that have a certain duration $T$. 
From universal scaling theory \cite{Sethna:2001} we expect the 
following power-law scaling relations in case of critical networks:
\begin{equation}
	f_{cml}(S) \sim S^{-\tau_{cml}}
\end{equation}
\begin{equation}
	f_{cml}(T) \sim T^{-\alpha_{cml}}
\end{equation}
\begin{equation}
	\langle S \rangle (T) \sim T^{1/ \sigma \nu z}
\end{equation}
where the exponents fulfill
\begin{equation}
	\frac{\alpha_{cml}}{\tau_{cml}} = \frac{1}{\sigma \nu z}.
\end{equation}

\section*{Results}

\subsection*{Adaptive network evolution}

In the following, we will have a look at different observables 
during numerical simulations of network evolution in the model. 
Key figures include the average connectivity $K$ 
and the branching parameter $\lambda$. Both are closely linked 
to, and influenced by, the ratio of activating links $p$ 
which is simply the fraction of positive couplings $c_{ij}=+1$ 
between all existing (non-zero) links.

\begin{figure}[!ht]
\begin{center}
\includegraphics[width=3.5in]{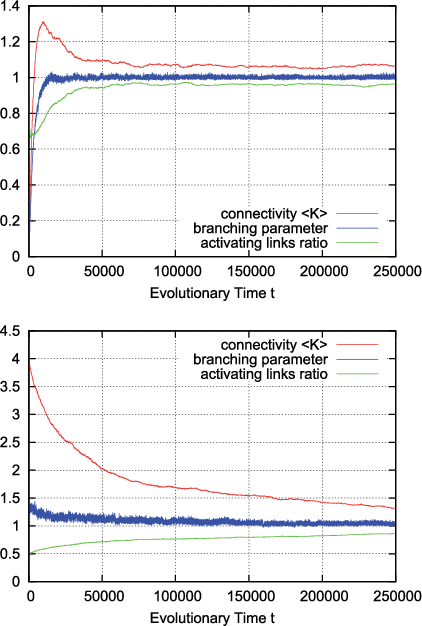}
\end{center}
\caption{
{\bf Typical run of the network self-organization algorithm.}  Regardless of initial connectivity and dynamical regime, the network evolves to a critical configuration. Top: when starting with completely isolated nodes the ``network'' is obviously subcritical and links will be inserted. Thus, both the connectivity (red) and the branching parameter (blue) as an indicator of network criticality increase. The network approaches a critical state where the branching parameter stabilizes close to one. Bottom: with higher initial connectivity, the network is supercritical at first. Links are removed from the network while the branching parameter approaches the critical value of one. As the self-organization algorithm is constructed to maximize activity correlations between linked nodes, the ratio of activating links (green) slowly increases in both cases.
}
\label{fig:01-evolution}
\end{figure}

The upper part in Figure \ref{fig:01-evolution} shows a typical  
run of the topology evolution algorithm, where we start with 
completely isolated nodes without any links. Trivially, the 
``network'' is subcritical at this stage, which can be seen 
from the branching parameter which is far below 1. 
As soon as rewiring takes place, the network starts to insert 
new links, obviously because these links enable the nodes to 
pass signals and subsequently act in a correlated way. 
With increasing connectivity, also the branching parameter rises, 
indicating that perturbations start to spread from their origin 
to other nodes. When the branching parameter approaches 1, 
indicating that the network reaches a critical state, 
the insertion of new links is cut back. The processes of 
insertion and depletion of links tend to be balanced 
against each other, regulating the network close to criticality.

On the other hand, if we start with a randomly interconnected 
network at a higher connectivity as, for example, $K_{ini}= 4$ 
(see lower part of Figure \ref{fig:01-evolution}), 
we find the network in the supercritical regime ($\lambda > 1$)
at the beginning. When above the critical threshold, 
many nodes will show chaotic activity with low average 
correlation to their respective inputs. The rewiring algorithm 
reacts by deleting links from the network, until the branching 
parameter approaches 1.
 
In both examples above, the evolution of the ratio of activating 
links $p$ (which tends towards 1) shows, that the rewiring algorithm 
in general favors the insertion of activating links and vice versa 
the deletion of inhibitory couplings. Indeed, this appears quite 
plausible when we remind ourselves that the rewiring mechanism 
optimizes the inputs of a node towards high correlation on average. 
Also, nodes will only switch to active state $\sigma_i=1$ if they 
get an overall positive input. As we had replaced spins by Boolean 
state nodes, this can only happen via activating links -- 
and that is why correlations mainly originate from positive 
couplings in our model. As a result, we observe the connectivity 
evolving towards one in-link per node, with the ratio of positive 
links also approaching one.

\begin{figure}[!ht]
\begin{center}
\includegraphics[width=3.5in]{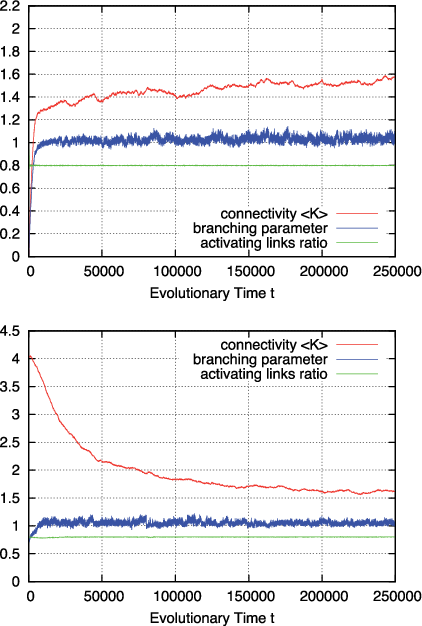}
\end{center}
\caption{
{\bf Typical run with fixed ratio of activating links.} If the ratio of activating links (green) is kept fixed (here: $p=0.8$; i.e.\ 80\% activating links) in order to keep some inhibiting links within the network, the connectivity (red) evolves to a higher value. Still, the branching parameter (green) is maintained close to the critical value of one. Top: starting with isolated nodes (subcritical). Bottom: starting at supercritical connectivity.
}
\label{fig:01-evolution-reshuffled}
\end{figure}

For a richer pattern complexity, we might later want to introduce 
a second mechanism which balances out positive and negative links, 
and with a first approach we can already test how the rewiring 
strategy would fit to that situation: if, after each rewiring step, 
we change the sign of single random links as necessary to obtain
a ratio of e.g.\ 80\% activating links (i.e.\ $p=0.8$), 
keeping the large majority of present links unchanged, 
the branching parameter will again stabilize close to the critical 
transition, while the connectivity is maintained at a higher value. 
Figure~\ref{fig:01-evolution-reshuffled} shows that the self-organization 
behavior is again independent from the initial conditions. 
This result does not depend on the exact choice of the activating 
links ratio $p$; similar plots can easily be obtained for a large 
range starting at $p = 0.5$, where the connectivity will subsequently 
evolve towards a value slightly below $K=2$, which is the critical 
connectivity for a randomly wired network with balanced link ratio 
according to the calculations made for the basic network model 
\cite{Rybarsch:2010}.

\subsection*{Avalanche properties}
\begin{figure}[!ht]
\begin{center}
\includegraphics[width=5.5in]{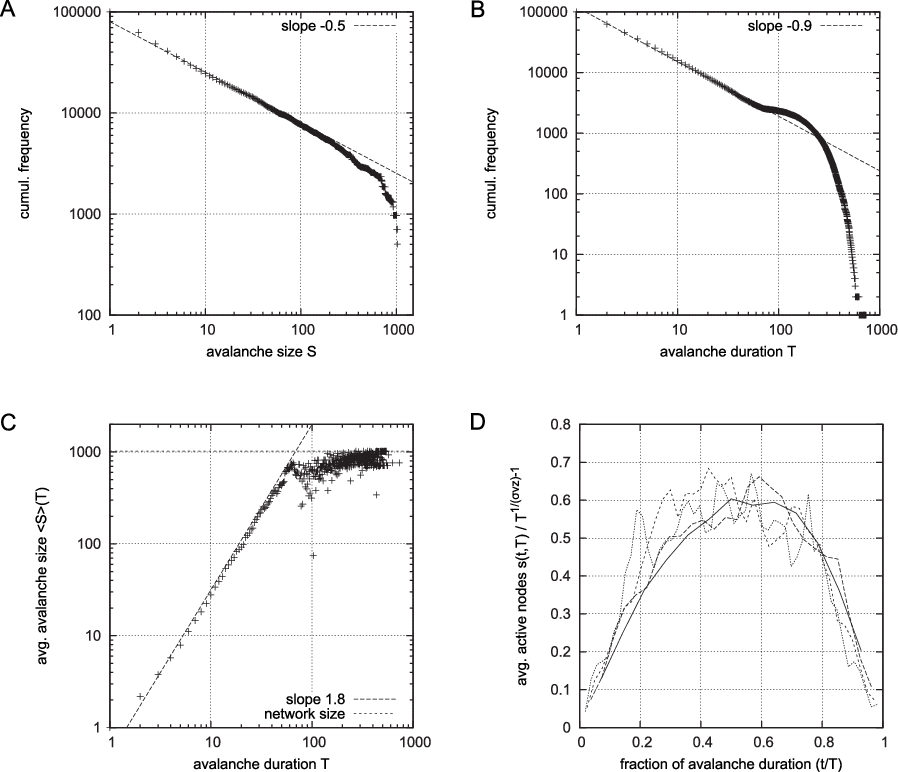}
\end{center}
\caption{
{\bf  Critical exponents and collapse of avalanche profiles.} Results measured from $10^5$ avalanches on 10 different evolved sample networks of $N=1024$ nodes, with an exhaust time parameter $\Omega = 1000$. Similar results were gained with $\Omega=100$ or $\Omega=10000$, this choice has no significant effect on the scaling exponents. A: Cumulative distribution of avalanche sizes shows a power-law scaling $f_{cml}(S) \sim S^{-\tau_{cml}}$ with exponent $\tau_{cml} \approx 0.5$. B: Cumulative distribution of avalanche sizes shows a power-law scaling $f_{cml}(T) \sim T^{-\alpha_{cml}}$ with exponent $\alpha_{cml} \approx 0.9$. C: Average size $\langle S \rangle$ of avalanches of duration $T$ shows a power-law increase corresponding to $\langle S \rangle (T) \sim T^{1/ \sigma \nu z}$ with an exponent of $\frac{1}{\sigma \nu z} \approx 1.8$. Note that the exponents $\tau_{cml}$, $\alpha_{cml}$, $\frac{1}{\sigma \nu z}$ fulfill the relation $\frac{\alpha_{cml}}{\tau_{cml}}=\frac{1}{\sigma \nu z}$ which is expected for a critical system. D: Profiles of avalanches, i.e. average scaled size as a function of scaled avalanche duration, of different durations (shown here for $T=13,26,39,52$) approximately collapse onto a universal shape.
}
\label{fig:01-avalanche-sizes-scaling}
\end{figure}

In addition to the branching parameter measurement, let us now  
take a look at some dynamical properties of the evolved networks 
to further characterize their criticality. 
Figure~\ref{fig:01-avalanche-sizes-scaling} shows the distributions of 
avalanche size and duration, as well as further scaling properties. 
(A): The avalanche size $S$ exhibits a power-law scaling 
\mbox{$f_{cml}(S) \sim S^{-\tau_{cml}}$} almost up to network 
size with an exponent \mbox{$\tau_{cml} = 0.5 \pm 0.05$} in the 
cumulative distribution, corresponding to a probability density 
exponent of \mbox{$\tau = 1.5 \pm 0.05$}. 
(B): Similarly, avalanche durations $T$ are power-law distributed 
as well up to a duration of approximately $70$ time steps, 
according to \mbox{$f_{cml}(T) \sim T^{-\alpha_{cml}}$} with 
an exponent of \mbox{$\alpha_{cml} = 0.9 \pm 0.05$}, 
i.e.\ \mbox{$\alpha = 1.9 \pm 0.05$}. As discussed above, 
plain power-law distributions can originate from several mechanisms 
and cannot be considered alone as clear evidence of criticality. 
To obtain a third exponent, we have also measured the average 
avalanche sizes $\langle S \rangle(T)$ as a function of avalanche 
duration $T$. It becomes clear from (C) that for durations of 
approximately $70$ time steps and more, the avalanches begin to 
span most of the system size, which explains the cutoff position 
in the avalanche duration scaling (B). Up to that point, 
we find a power-law scaling 
\mbox{$\langle S \rangle (T) \sim T^{1/ \sigma \nu z}$} 
with an exponent of \mbox{$\frac{1}{\sigma \nu z} = 1.8 \pm 0.05$}. 
These exponents are both in line with experimental results 
\cite{Friedman:2012} and fulfill the exponent relation 
\mbox{$\frac{\alpha_{cml}}{\tau_{cml}}= \frac{\alpha -1}{\tau -1} 
= \frac{1}{\sigma \nu z}$} as predicted for a critical system 
by the scaling theory of non-equilibrium critical phenomena \cite{Sethna:2001}. 
(D): Finally, we find that avalanche profiles (i.e. average scaled 
size as a function of scaled avalanche duration) of avalanches of 
different durations $T$ approximately collapse onto a universal shape, 
another feature of criticality also seen in the neural experiments 
\cite{Friedman:2012}.  

\subsection*{Variations in activation thresholds and response to external perturbation}

\begin{figure}[!ht]
\begin{center}
\includegraphics[width=3.5in]{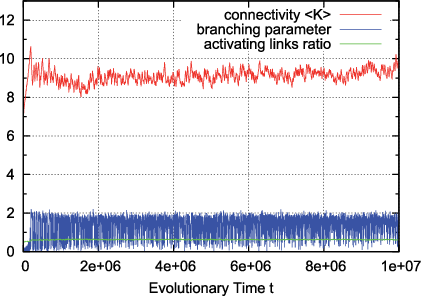}
\end{center}
\caption{
{\bf Typical run with higher activation thresholds \mbox{$\Theta_i = 1$}.} When activation thresholds are increased, a node needs more than one excitatory input to become active itself. Thus, higher overall connectivity is needed to allow signal propagation on a critical level. The adaptation process responds accordingly and maintains a connectivity around $K=9$ while the branching parameter shows larger fluctuations between sub- and supercritical states, but in general is kept on a moderate level and does not diverge with increasing connectivity.
}
\label{fig:01-evolution-theta1}
\end{figure}

In the above simulations, the activation thresholds of all nodes 
were strictly set to \mbox{$\Theta_i = 0$} for maximum model simplicity. 
However, a neuron might as well need higher input to become active. 
Figure~\ref{fig:01-evolution-theta1} demonstrates that the proposed 
adaptation algorithm similarly works well on networks of nodes with 
a non-zero activation threshold of e.g.\ \mbox{$\Theta_i = 1$}. 
According to the update rule \eqref{eq:01-update}, now at least two 
positive inputs are necessary to activate a single node. 
As the rewiring algorithm is based on propagation of thermal noise signals, 
the inverse temperature $\beta$ needs to be selected at a lower 
value than before. (As a general rule, $\beta$ should be selected 
in a range where thermal activation of nodes occurs at a low rate, 
such that avalanches can be triggered, but are not dominated by noise.) 
The simulation is now started at an average connectivity of $K=7$ 
which is still sub-critical in this case (branching parameter low). 
In a similar way as shown above, the network adapts by inserting 
new links and increasing $K$, thereby also increasing the average 
branching parameter. While the system does not 
approach to a phase transition as nicely as shown above for 
activation thresholds of zero (in fact the branching fluctuates 
much more around the target value of one), the general tendency 
remains: the rewiring mechanism reacts properly before the network 
drifts too far off from criticality. The connectivity also fluctuates 
more, but stabilizes on a level around $K=9$.

In their \emph{in-vitro} experiments, Beggs and Plenz further 
demonstrate that cortical networks can self-regulate in response 
to external influences, retaining their functionality while avalanche-like 
dynamics persist -- for example after neuronal excitability has been 
increased by adding stimulant substances to the cultures.

To reproduce such behavior in our model, we can include variations 
in the activation thresholds $\Theta_i$ of the individual nodes. 
Assume we start our network evolution algorithm with a moderately 
connected, but subcritical network, where all nodes have an activation 
threshold of $\Theta_i = 1$. 

\begin{figure}[!ht]
\begin{center}
\includegraphics[width=3.5in]{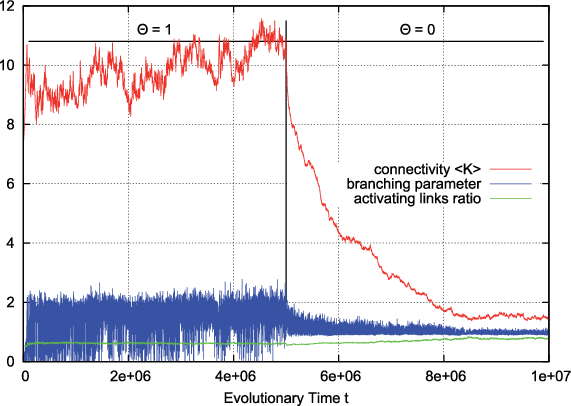}
\end{center}
\caption{
{\bf Rewiring response to a sudden decrease of activation thresholds.} If we first set all node activation thresholds to $\Theta_i = 1$, the connectivity (red) must first evolve to a higher value to allow propagation of activity within the network. When all activation thresholds are suddenly reduced (to mimic an external influence of neuronal excitability by stimulant substances) at the same time step, the network properly responds to the new situation and reduces connectivity to decrease excitability back to a critical level.
}
\label{fig:01-variable-theta} 
\end{figure}

Figure~\ref{fig:01-variable-theta} shows that the network first behaves 
in the same way as demonstrated in Figure~\ref{fig:01-evolution-theta1} 
for activation thresholds $\Theta_i = 1$. At one time step in the center 
of Figure~\ref{fig:01-variable-theta}, we at once reset all nodes to 
an activation threshold of $\Theta_i = 0$, simulating the addition 
of a stimulant. As we can expect, this immediately puts the network 
into a supercritical, chaotic state. This is reflected by the branching 
parameter, which now constantly stays above one and does not fluctuate 
below anymore. It is clearly visible that the rewiring mechanism 
promptly reacts and drives back the connectivity, until the branching 
parameter slowly converges towards one again. A similar behavior is 
also found if thresholds $\Theta_i$ are not changed all at once, 
but gradually during network evolution.

\subsection*{Summary and Discussion} 
To conclude, we have demonstrated that a very minimalistic binary 
neural network model is  capable of self-organized critical behavior 
that matches the experimentally observed criticality in neural systems. 

We revisited the earliest spin model for self-organized critical 
neural networks and transformed it to networks of nodes with 
Boolean node states and with arbitrary topology. The adaptive 
dynamics of the network is a simple, locally realizable rewiring 
mechanism which uses activity correlation as its regulation criterion, 
thus retaining the biologically inspired rewiring basis from the 
spin version of the original model. As a result the dynamical network 
exhibits emerging activity avalanches with spatio-temporal properties 
comparable to those observed in real neuronal systems.  

What the model does not provide are hypotheses about possible details 
of implementations on the biological level. We did not make particular 
efforts to implement a fully local version, although such local, 
continuously running versions of the algorithm are straightforward. 
Instead we kept the stepwise procedure of separate correlation 
measurements at two different times for clarity. A biological 
implementation, in one form or another, has to sense the time 
derivative of the correlation for which there are numerous 
possibilities. Apart from that central detail it is obvious that 
on the local level there are far more details possible in a 
biological realization -- some of which are contained in other 
existing models reviewed above -- which we do not further discuss here. 
However, the central properties of criticality will be independent 
of these details. For future work, it might be fruitful to study 
particular biological realizations of the correlation 
based adaptation which we here studied in a bare bones 
algorithmic version. Further, it will be interesting to compare
our algorithm with certain other models for neural adaptation
with particular attention to the scaling properties at criticality. 

In summary we find that the earliest spin model of neural criticality 
exhibits avalanche statistics that compare
well with experimental data without the need for parameter tuning.
The model represents a fundamental organization mechanism 
leading to a critical system that may serve as the simplest 
representative of a ``neural SOC universality class'', matching the 
observed characteristics of self-organized criticality in biological 
cortical tissue. 
In particular, the model exhibits a scaling of avalanche size 
and duration distributions, as well as a universal scaling of 
the temporal avalanche profiles, altogether constituting  
the specific characteristics of neuronal avalanches near criticality.

\end{document}